\documentclass[twocolumn,showpacs,amsmath,amssymb,10pt]{revtex4}
\usepackage{bm}
\usepackage{amsmath,amsfonts,latexsym,graphicx,amssymb}

\newcommand{\ot}{{\,\otimes\,}}
\newcommand{{\Cd}}{{\mathbb{C}^d}}

\def\<{\langle}
\def\>{\rangle}

\newtheorem{proposition}{Proposition}

\newtheorem{lemma}{Lemma}

\begin{document}

\title{\textbf{Disproving  the conjecture on structural physical approximation \\ to optimal decomposable entanglement witnesses
%Witnessing entanglement in $4N \times 4N$ systems
}} \author{Dariusz
Chru\'sci\'nski and Gniewomir Sarbicki\thanks{email:
darch@fizyka.umk.pl}}
\affiliation{Institute of Physics, Nicolaus Copernicus University,\\
Grudzi\c{a}dzka 5/7, 87--100 Toru\'n, Poland}

%\begin{document}

\begin{abstract}
We disprove the conjecture that
structural physical approximations to optimal entanglement witnesses are separable states. The conjecture holds true for extremal decomposable entanglement witnesses.

\end{abstract}
\pacs{03.65.Ud, 03.67.-a}

\maketitle

Few years ago authors of \cite{SPA} posed the following conjecture: Structural Physical Approximations (SPA) to optimal positive
maps are entanglement breaking channels. In the language of entanglement witnesses this conjecture may be reformulated as follows: Let $W$ be a normalized (i.e. ${\rm Tr}\, W=1$) entanglement witness (EW) acting on a Hilbert space $\mathcal{H}=\mathcal{H}_A \ot \mathcal{H}_B$ of finite dimension $D=d_A \times d_B$. Consider the following convex combination
\begin{equation}\label{}
  W(p)  = pW + \frac{1-p}{D}\, \mathbb{I}\ ,
\end{equation}
i.e. a mixture of $W$ with a maximally mixed state $\mathbb{I}/D$. It is clear that there exists the largest $p_* \in (0,1)$ such that for all $p \leq p_*$ the operator $W(p) \geq 0$ and hence it defines a legitimate quantum states. One calls $W(p_*)$ the Structural Physical Approximations of $W$. SPA conjecture \cite{SPA} states that if $W$ is an optimal EW, then its SPA $W(p_*)$ defines a separable state. It is clear that if $W(p_*)$ is separable, then for all $p\leq p_*$ a state $W(p)$ is separable as well. This conjecture was supported by numerous examples of witnesses (see, e.g., Ref. \cite{DC,A}), both
decomposable and indecomposable, and also analyzed in the
continuous variables case \cite{A}.

Recently SPA conjecture has been
disproved by Ha and Kye \cite{Ha}, who have provided an example
of an indencomposable optimal entanglement witness, whose
structural physical approximation is entangled. The conjecture remains still open in the decomposable case (see recent discussion in \cite{D1,D2,Chiny}). In the present paper we show that
%
%\begin{itemize}
%
%\item 
the SPA approximation to optimal decomposable witness may provide an entangled state.
%
%\item the SPA approximation to extremal decomposable witness always provides a separable state.
%
%\end{itemize} 
Let us recall that a Hermitian  operator $W$ acting in $\mathcal{H}_A \ot \mathcal{H}_B$ is an EW iff $i)$ $\< \psi \ot \phi |W|\psi \ot \phi\> \geq 0$, and $ii)$ $W \ngeq 0$, i.e. $W$ possesses at least one negative eigenvalue.
The optimal witness is defined as follows: if $W_1$ and $W_2$ are two entanglement witnesses then following Ref. \cite{OPT} we call $W_1$ finer than $W_2$ if $D_{W_1} \supseteq  D_{W_2}$, where
\begin{equation}\label{}
  D_{W} = \{ \, \rho\, |\, {\rm Tr}(\rho W) < 0 \, \}\ ,
\end{equation}
denotes the set of all entangled states detected by $W$. Now,  an EW $W$ is optimal if there is no other witness
that is finer than $W$. One proves \cite{OPT} that $W$ is optimal iff for any $\alpha > 0$ and a positive operator $P$ an operator $W - \alpha P$ is no longer an EW. Authors of \cite{OPT} provided the following sufficient condition of optimality: for a give EW $W$ define
\begin{equation*}\label{}
  P_W = \{\, |\psi \ot \phi\> \in \mathcal{H}_A \ot \mathcal{H}_B\, |\, \< \psi \ot \phi |W|\psi \ot \phi\>  = 0 \, \} \ .
\end{equation*}
Now, if $P_W$ spans $\mathcal{H}_A \ot \mathcal{H}_B$, then $W$ is optimal. Entanglement witness $W$ is decomposable (DEW) iff
\begin{equation}\label{}
  W = A + B^\Gamma\ ,
\end{equation}
where $A,B \geq 0$ and $B^\Gamma$ denotes partial transposition of $B$. Clearly, if DEW $W$ is optimal, then  $W=B^\Gamma$ and  $B$  supported on a completely entangled subspace (CES) of $\mathcal{H}$ (a linear subspace $\Sigma \subset \mathcal{H}$ is called  CES if there is no non-zero product vectors in $\Sigma$).
Now, if DEW $W$ is optimal, then
\begin{equation}\label{}
  W(p_*)^\Gamma = p_* W + \frac{1-p_*}{D} \mathbb{I} \ ,
\end{equation}
which shows that SPA for an optimal DEW is a PPT state (a state $\rho$ is PPT if $\rho^\Gamma \geq 0$). SPA conjecture states that $W(p_*)$ is not only PPT but also separable. Since in $2 \ot 2$ and $2\ot 3$ all PPT states are separable, the SPA conjecture is trivially satisfied for qubit-qubit and qubit-qutrit systems. In general case separable states define only a proper subset of PPT states  and hence one may have PPT states which are entangled.

Now, we show that the SPA conjecture for optimal DEWs is not true. Consider the following family of bipartite Hermitian operators in $\mathbb{C}^3 \ot \mathbb{C}^3$
\begin{equation}\label{W-gamma}
  W_\gamma = 3 B_\gamma^\Gamma\ ,
\end{equation}
with
\begin{equation}
  B_{\gamma} =  \frac{1-\gamma}2 P_{10} + \frac{1-\gamma}2 P_{20} + \gamma P_{11}\ ,
 \end{equation}
where $P_{kl}=|\Omega_{kl}\>\<\Omega_{kl}|$ denotes a set of rank-1 projectors with $|\Omega_{kl}\> = \mathbb{I} \ot W_{kl} |\Omega_{00}\>$ and $W_{kl}$ is a Weyl operator defined by $W_{kl}|i\> = \omega^{k(i-l)}|i-l\>$ with $\omega = e^{2\pi i}/{3}$. Finally, $|\Omega_{00}\> = \frac{1}{\sqrt{3}}\sum_{i=0}^2 |ii\>$ denotes a maximally entangled state of two qutrits and $\gamma$ is a real parameter. Hence $B_\gamma$ belongs to a class of Bell diagonal operators considered in \cite{Wieden}. One finds
\begin{eqnarray*}
  |\Omega_{10}\> &=& \frac{1}{\sqrt{3}} ( |00\> + \omega |11\> + \omega^* |22\> ) \ ,\\
  |\Omega_{20}\> &=& \frac{1}{\sqrt{3}} ( |00\> + \omega^* |11\> + \omega |22\> ) \ ,\\
  |\Omega_{11}\> &=& \frac{1}{\sqrt{3}} ( \omega^* |02\> +  |10\> + \omega |21\> ) \ .
\end{eqnarray*}
\begin{proposition}
For each $\gamma \in (0,1)$ an operator $W_\gamma$ is an optimal entanglement witness.
\end{proposition}
Proof: let us consider a matrix form of $W_\gamma$
\begin{widetext}
 \begin{equation}
  W_{\gamma} =
  \left[ \begin{array}{ccc|ccc|ccc}
  1-\gamma & \cdot & \cdot & \cdot & \cdot & \cdot & \cdot & \omega \gamma & \cdot \\
  \cdot & \gamma & \cdot & -\frac{1-\gamma}2 & \cdot & \cdot & \cdot & \cdot & \cdot \\
  \cdot & \cdot & \cdot & \cdot & \omega^* \gamma & \cdot & -\frac{1-\gamma}2 & \cdot & \cdot \\
  \hline
  \cdot & -\frac{1-\gamma}2 & \cdot & \cdot & \cdot & \cdot & \cdot & \cdot & \omega^*\gamma \\
  \cdot & \cdot & \omega\gamma & \cdot & 1-\gamma & \cdot & \cdot & \cdot & \cdot \\
  \cdot & \cdot & \cdot & \cdot & \cdot & \gamma & \cdot & -\frac{1-\gamma}2 & \cdot \\
  \hline
  \cdot & \cdot & -\frac{1-\gamma}2 & \cdot & \cdot & \cdot & \gamma & \cdot & \cdot \\
  \omega^*\gamma & \cdot & \cdot & \cdot & \cdot & -\frac{1-\gamma}2 & \cdot & \cdot & \cdot \\
  \cdot & \cdot & \cdot & \omega \gamma & \cdot & \cdot & \cdot & \cdot & 1-\gamma \\
  \end{array} \right]\ ,
 \end{equation}
\end{widetext}
where, to make the picture more transparent we replace zeros by dots. To prove that $W_\gamma$ is an entanglement witness it is enough to show that it has negative eigenvalue. Note that the following $3 \times 3$ sub-matrix
 \begin{equation*}
 \left[\begin{array}{ccc}1-\gamma & 0 & \omega \gamma \\ 0 & \gamma & -\frac{1-\gamma}2 \\\
  \omega^* \gamma & -\frac{1-\gamma}2 & 0
\end{array}\right]
 \end{equation*}
has exactly one negative eigenvalue. Actually,  $W_\gamma $ has exactly one negative eigenvalue with 3-fold degeneracy. To show that $W_\gamma$ is optimal
we prove that the corresponding set $P_{W_\gamma}$ spans $\mathbb{C}^3 \ot \mathbb{C}^3$ \cite{OPT}. Now, if $W=B^\Gamma$, then $\< \psi \ot \phi|W| \psi \ot \phi\> = \<\psi \ot \phi^*|B|\psi \ot \phi^*\>$. Let us look for vectors $x,y \in \mathbb{C}^3$ satisfying
\begin{equation}\label{Pxy}
  \< x \ot y | B_\gamma |x \ot y \> =0\ .
\end{equation}
Note that for $\gamma \in (0,1)$ one has $B_\gamma \geq 0$ and hence it is clear that any such vector $|x \ot y\>$ has to be orthogonal to $|\Omega_{10}\>$,  $|\Omega_{20}\>$ and $|\Omega_{11}\>$. Orthogonality to $|\Omega_{10}\>$ and   $|\Omega_{20}\>$ gives
\begin{eqnarray*}
  a_0 + \omega a_1 + \omega^* a_2 &=& 0\ , \\
  a_0 + \omega^* a_1 + \omega a_2 &=& 0\ ,
\end{eqnarray*}
where $a_k = x_k y_k$. These equations imply $a_0=a_1=a_2$. Since the norm of $|x\ot y\>$ does nor play any role we assume $a_k=1$, that is $y_k = 1/x_k$ for $k=0,1,2$. Finally, orthogonality to $|\Omega_{11}\>$ implies
\begin{equation}\label{!}
  \frac{x_1}{x_2} + \omega \frac{x_2}{x_3} + \omega^* \frac{x_3}{x_1} = 0 \ .
\end{equation}
Hence product vectors $|x \ot y\> = \sum_{i,j=0}^2 ({x_i}/{x_j}) \, |ij\>$, where $x_i$ are constrained by (\ref{!}), satisfy   (\ref{Pxy}). It is clear that these vectors span 6-dimensional subspace in $\mathbb{C}^3 \ot \mathbb{C}^3$. Now, $|x \ot y^*\> = \sum_{i,j=0}^2 ({x_i}/{x_j^*}) \, |ij\>$ together with (\ref{!}), satisfy $\< x\ot y^*|W_\gamma|x\ot y^*\> =0$. Note that functions $x_i/x_j^*$ are linearly independent and eq. (\ref{!}) does not introduce any additional constraint among them. Hence, vectors $|x\ot y^*\>$ span the entire 9-dimensional Hilbert space $\mathbb{C}^3 \ot \mathbb{C}^3$ which proves optimality of $W_\gamma$. \hfill $\Box$

As a byproduct we showed that 3-dimensional subspace spanned by $\{ |\Omega_{10}\>,|\Omega_{20}\>,|\Omega_{11}\> \}$ defines CES in $\mathbb{C}^3 \ot \mathbb{C}^3$. Now, let us consider the SPA for $W_\gamma$. To check for separability of $W_\gamma(p_*)$ we apply well known realignment criterion \cite{REL}. Numerical analysis shows that at least for some range of $\gamma$ the state $W_\gamma(p_*)$ is entangled. Fig. 1 shows the plot of $R(W_\gamma(p_*)) - 1$. Recall, that if $||R(\rho)||_1 > 1$, then $\rho$ is entangled.

%------------------------
\begin{figure}[h]
\begin{center}
\includegraphics[width=0.8\columnwidth]{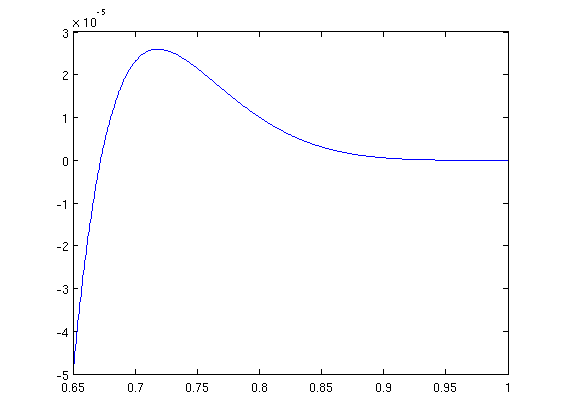}
\end{center}
\caption{\label{F} The plot of  $||R(W_\gamma(p_*)||_1  - 1$.}
\end{figure}
%------------------------
Actually, in this case one can compute the realignment analytically (see the Appendix).

This way the SPA conjecture is finally disproved. Interestingly, as was already shown in \cite{SPA}, the conjecture is true for a proper subset of optimal decomposable entanglement witnesses, i.e. witnesses of the of the form  $W = P^\Gamma$ and $P=|\psi\>\<\psi|$ with entangled $|\psi\>$.

%Hence, our basic question is whether a PPT state $W(p_*)$ is separable. We show that the answer is positive if $W$ is not only optimal but also extremal:

\vspace{.3cm}

\noindent To summarize: we have found a counterexample disproving  SPA conjecture for optimal decomposable entanglement witnesses. The idea to use the family $W_\gamma$ comes from \cite{Wieden} where the Bell diagonal states with bound entanglement were considered.

\acknowledgements
This work was partially supported by the National Science Center project  %{\em Quantum correlations: analysis, detection and dynamics}.
DEC-2011/03/B/ST2/00136.

\appendix

\section*{Appendix}

We show analytically that for $\gamma=3/4$ the PPT state $W_\gamma(p_*)$ is entangled.
Let $\lambda_- <0$ be the smallest eigenvalue of $W_\gamma$ (recall, that there is only one 3-fold degenerate negative eigenvalue) and let $Q_\gamma = W_\gamma - \lambda_- \mathbb{I}$. We calculate $||R(Q_\gamma)||_1 $ and compare with ${\rm Tr}\, Q_\gamma$.  One finds for $R(Q_\gamma)R(Q_\gamma)^\dagger$:
%\begin{widetext}
\begin{equation*} \label{real^2}
% R(\rho_\gamma)R(\rho_\gamma)^\dagger =
  \left[ \begin{array}{ccc|ccc|ccc}
 d_1 & \cdot & \cdot & \cdot & q_1 & \cdot & \cdot & \cdot & q_1 \\
 \cdot & d_2 & \cdot & \cdot & \cdot & q_2\omega^* & q_2\omega & \cdot & \cdot \\
 \cdot & \cdot &  d_2 & q_2\omega & \cdot & \cdot & \cdot & q_2\omega^* & \cdot \\ \hline
 \cdot & \cdot & q_2\omega^* & d_2 & \cdot & \cdot & \cdot & q_2\omega & \cdot \\
 q_1 & \cdot & \cdot & \cdot & d_1 & \cdot & \cdot & \cdot & q_1 \\
 \cdot & q_2\omega & \cdot & \cdot & \cdot & d_2 & q_2\omega^* & \cdot & \cdot \\ \hline
 \cdot & q_2\omega^* & \cdot & \cdot & \cdot & q_2\omega & d_2 & \cdot & \cdot \\
 \cdot & \cdot & q_2\omega & q_2\omega^* & \cdot & \cdot & \cdot & d_2 & \cdot \\
 q_1 & \cdot & \cdot & \cdot & q_1 & \cdot & \cdot & \cdot & d_1
 \end{array} \right],
\end{equation*}
%\end{widetext}
 where
\begin{eqnarray*}
  d_1 &=& (\gamma + \lambda_- - 1)^2 + \lambda_-^2 + (\gamma - \lambda_-)^2\ , \\
  d_2 &=& \gamma^2 + \left(\frac{\gamma-1}2\right)^2\ , \\
  q_1 &=& \lambda_-(\gamma + \lambda_- - 1)  - (\gamma - \lambda_-)(\gamma + 2\lambda_- - 1) \ ,  \\
  q_2 &=& \gamma\frac{\gamma-1}2 \ .
\end{eqnarray*}
Note that  $R(Q_\gamma)R(Q_\gamma)^\dagger$ is a direct sum of three $3\times 3$ matrices:
\begin{displaymath}
A_1 = \left[ \begin{array}{ccc} d_1 & q_1 & q_1 \\ q_1 & d_1 & d_1 \\ d_1 & d_1 & d_1 \end{array} \right] \ , \quad
A_2 =\left[ \begin{array}{ccc} d_2 & q_2\omega^* & q_2\omega \\ d_2\omega & d_2 & q_2\omega^* \\ q_2\omega^* & q_2\omega & d_2 \end{array} \right]\ , % \left[  % \begin{array}{ccc} D_2 & P_2\omega & P_2\omega^* \\ P_2\omega^* & D_2 & P_2\omega \\ P_2\omega & P_2\omega^* & D_2 \end{array} \right]
 \end{displaymath}
 and
$$ A_3 = \left[ \begin{array}{ccc} d_2 & q_2\omega & q_2\omega^* \\ q_2\omega^* & d_2 & q_2\omega \\ q_2\omega & q_2\omega^* & d_2 \end{array} \right] \ . $$
Note that $A_2$ and $A_3$ are similar and hence have the same spectrum. The eigenvalues of $A_1$ read:
single   $d_1+2q_1 = (1-3\lambda_-)^2$ and double $d_1-q_1 = 3\gamma^2-3\gamma+1$. Similarly the eigenvalues of $A_1$ read:
single $d_2+2q_2 = \frac 14 (3\gamma-1)^2$ and double  $d_2-q_2 = \frac 14 (3\gamma^2+1)$. Hence
\begin{eqnarray*}\label{}
||R(Q_\gamma)||_1 &=& 3\gamma-1 + 1 - 3\lambda_- + 2\sqrt{3\gamma^2-3\gamma+1} \nonumber \\ &+& 2\sqrt{3\gamma^2+1} \ .
\end{eqnarray*}
Now, according to the realignment criterion $Q_\gamma$ is entangled if $||R(Q_\gamma)||_1 > {\rm Tr}\, Q_\gamma$, that is, $\lambda_- > \lambda_0$, where
\begin{displaymath}
  \lambda_0 = \frac{1-\gamma}2 - \frac 13 \sqrt{3\gamma^2-3\gamma+1} -\frac 13 \sqrt{3\gamma^2+1} .
\end{displaymath}
For $\gamma=\frac 34$ the characteristic polynomial of $W_{3/4}$ is equal $P(\lambda)=-\lambda^3+\lambda^2+\frac{25}{64}\lambda-\frac{109}{256}$ and $\lambda_0$ is equal $\frac 18 (1-\frac 23 \sqrt{7}-\frac 23 \sqrt{43}) \approx -0.64193$. Take a slightly bigger number $\lambda'= -0.64191> \lambda_0$. One finds $P(\lambda') \approx 1.91 \times 10^{-5}> 0$ and because there is only one 3-fold degenerate negative eigenvalue of $W_{3/4}$, the number $\lambda'$ satisfies $\lambda'< \lambda_-$ and hence $\lambda_- > \lambda_0$ which proves that the realignment of $\rho_{3/4} = Q_{3/4}/{\rm Tr} Q_{3/4}$ is strictly greater than 1 which implies that $\rho_{3/4}$ is an entangled PPT state.


\newpage

\begin{thebibliography}{1} \bibliographystyle{plain}

\bibitem{SPA} J.K. Korbicz, M.L. Almeida, J. Bae, M. Lewenstein and
A. Acin, Phys. Rev. A {\bf 78}, 062105 (2008).

\bibitem{DC} D. Chru\'sci\'nski and J. Pytel, Phys. Rev. A {\bf 82}, 052310 (2010);
D. Chru\'sci\'nski, J. Pytel, G. Sarbicki, ibid. {\bf 80}, 062314 (2009).

\bibitem{A}  R. Augusiak, J. Bae, {\L}. Czekaj, M. Lewenstein, J. Phys. A {\bf 44},
185308 (2011).


\bibitem{Ha} K.-C. Ha and S.-H. Kye, J. Math. Phys. {\bf 53}, 102204 (2012).


\bibitem{D1} K.-C. Ha and S.-H. Kye, {\em Separable states with unique decompositions},
arXiv:1210.1088.

\bibitem{D2} R. Augusiak, J. Bae, J. Tura, and M. Lewenstein, {\em Comment on “Separable states with unique decompositions”}, arXiv.1304.2040.

\bibitem{Chiny} B.-H. Wang and D.-Y. Long, Phys. Rev. A {\bf 87}, 062324 (2013).

\bibitem{OPT} M. Lewenstein, B. Kraus, J. I. Cirac, and P. Horodecki, Phys.
Rev. A {\bf 62}, 052310 (2000).

\bibitem{Wieden} B. Baumgartner, B.C. Hiesmayr, and H. Narnhofer, Phys. Lett. A {\bf 372}, 2190 (2008).

\bibitem{REL}  K. Chen and L.A. Wu, Quantum Inf. Comput. {\bf 3}, 193 (2003).

%\bibitem{Vidal} G. Vidal and R. Tarrach, Phys. Rev. A {\bf 59}, 141
%(1999).



\end{thebibliography}
\end{document}